\documentclass[twocolumn,showpacs,pra,amsmath,amssymb,floatfix]{revtex4-1}
\input{epsf}

\usepackage{graphicx}
\usepackage{dcolumn}
\usepackage{bm}

\begin{document}
\title{Topological phases in spin-orbit coupled 
dipolar lattice bosons}
\author{H. T. Ng}
\affiliation{Center for Quantum Information, Institute for Interdisciplinary Information Sciences, Tsinghua University, Beijing 100084, P. R. China}

\date{\today}

\begin{abstract}
We study the topological phases in spin-orbit coupled dipolar bosons in a one-dimensional optical lattice. 
The magnetic dipolar interactions between atoms give rise to the inter-site interactions.
In the Mott-insulating regime, this system can
be described by the quantum XYZ spin model with the Dzyaloshinskii-Moriya interactions in a transverse field.
We focus on investigating the effect of dipolar interactions on the topological phase.
The topological phase can be shown when spin-orbit coupling
incorporates with the repulsive dipolar interaction. 
We find that the dipolar 
interaction can broaden the range of parameters of spin-orbit coupling
and transverse field for exhibiting the topological 
phase. The sum of spin correlations between the two nearest neighbouring atoms can be used to indicate the topological phase. This may be useful for detecting topological phases in experiments.
\end{abstract}


\maketitle

\section{Introduction}
Majorana fermions are exotic
particles, where their anti-particles
are their own particles \cite{Wilczek}. 
Since Kitaev \cite{Kitaev} found
that an unbound pair of Majorana
fermions can be realized in a
one-dimensional (1D) spin-polarized
superconductor, a considerable 
effort has been devoted to studying
semiconducting wires \cite{Lutchyn,Oreg} and cold atoms in
optical lattices \cite{Jiang,Kraus} for realization of Majorana
fermions. In addition, Majorana fermions are 
robust against local perturbations 
due to the topological degeneracy.
Therefore, it may be potentially 
applied to quantum information
processing such as quantum memory \cite{Mazza}
and topological quantum computing \cite{Nayak,Alicea}.

Ultracold atoms have been used for
simulating various quantum many-body
phenomena \cite{Bloch}. For example, 
quantum phase transition from a superfluid
to a Mott insulator \cite{Greiner} has been shown
in lattice bosons. Also,
spin-orbit (SO) coupling in a Bose-Einstein
condensate (BEC) has been shown
in an experiment \cite{Lin}, where the two
atomic spin states are coupled to their
momentum states by using a pair of lasers. 
In fact, SO coupling 
is essential to realize topological matter 
\cite{Hasan,Qi}. Therefore, ultracold atoms  
pave the way for studying topological phases \cite{Galitski}.

Recently, SO-coupled dipolar condensates
have been theoretically studied \cite{Deng,Wilson}. 
The intriguing ground state of ultracold atoms 
can be displayed via
the interplay of SO coupling and magnetic 
dipole-dipole interaction (DDI). Indeed, dipolar interactions
in ultracold atomic gases can give rise to 
spectacular quantum phenomena \cite{Lahaye} such as 
spin texture \cite{Vengalattore} and Einstein-de Haas effect \cite{Swislocki}.
In addition, a chromium gas in an
optical lattice \cite{Paz0,Paz} has recently been realized.
Since a chromium atom has a relatively large magnetic dipole
moment \cite{Santos}, the inter-site interactions become
strong enough \cite{Paz} to experimentally study 
quantum magnetism \cite{Auerbach}.

In this paper, we consider 
spin-orbit coupled bosonic atoms in a 1D
optical lattice, where the atoms at
the different sites couple to each 
other via the magnetic dipolar 
interactions \cite{Paz}. In the Mott-insulating
regime, this system can be described
by the quantum XYZ spin model with the Dzyaloshinskii-Moriya (DM) interactions in a transverse
field \cite{Zhou,Piraud,Zhao}, where a two-component 
boson in each site
can be viewed as a spin-half particle \cite{Duan}.
Indeed, a spin chain is equivalent
to a 1D spinless fermion model by performing
the Jordan-Wigner transformation \cite{Kitaev2}. 
This enables us to study the topological
phenomenon, such as Majorana fermions, with a quantum spin chain \cite{Kitaev2,Mezzacapo}.

The long-ranged DDI leads to interactions between
the two atoms at the different sites.
It is important to examine the effect
of long-ranged interactions on the topological
phases \cite{Stoudenmire,Gangadharaiah,Sela,Hassler,Thomale}. 
Although the magnetic dipolar interactions
between the atoms are weak, such strengths
are comparable to effective coupling strengths
between the spins \cite{Paz} in the Mott-insulating
regime. Therefore, {\it the effect of SO coupling, DDI 
and transverse field on the topological
phases can be studied in a controllable manner}.

Here we study the energy difference
between two degenerate ground states in the two
different parity sectors  
to determine the topological phase \cite{Stoudenmire}. 
When the energy
difference becomes zero, the topological
degeneracy occurs in the system.
Although the DM interactions break the 
$\mathbb{Z}_2$ symmetry in this system \cite{Kitaev2}, the system
exhibits the topological phase if the 
DM interactions are sufficiently small.
We find that the SO couplingz incorporate with 
repulsive DDIz can be used to 
exhibit topological phases.
A topological phase is shown when
the SO coupling strength increases
with the dipolar interaction strength.
Also, the DDI can broaden the range 
of SO coupling and transverse field strengths for 
exhibiting topological phases. 
Therefore, this may lead to observing topological phases
in 1D SO-coupled dipolar lattice bosons.

In addition, we show that the sum of all 
spin-polarization correlations 
between nearest neighbouring atoms can be
used for characterizing the magnetic
properties of a spin chain.
When the system is in a topological
phase, the sum of spin correlations 
is close to zero.
This coincides with the topological 
phase which is determined by using
the energy gap between the two different 
parity sectors. In fact, spin 
correlations between the spins
in the different sites have been probed in a 
1D optical lattice \cite{Endres}.
Thus, this can be used for detecting
the topological phases in experiments. 

This paper is organized as follows:
In Sec.~II, we introduce the system
of SO-coupled dipolar bosons in a 1D
lattice. In Sec.~III, we discuss
the system in the Mott-insulating regime.
In Sec.~IV, we study the effect of
dipolar interactions and the transverse field
on topological phases.  
We provide a summary in Sec.~V.

\section{System}
We consider the two-component dipolar bosonic atoms to be
trapped a 1D optical lattice, where the atoms are spin-orbit coupled \cite{Lin}. Recently, a spin-orbit coupled dipolar condensate
has recently been discussed \cite{Deng,Wilson}, where the two internal states
can be chosen in the ground electronic manifold \cite{Deng} for
SO coupling.    
In this system, the atoms interact with each other
via short-range and long-range interactions, and
the two neighbouring atoms are coupled
through SO coupling.
The total system can
be described by a Hamiltonian:
\begin{eqnarray}
H&=&H_{\rm BH}+H_{\rm SO}+H_{d},
\end{eqnarray}
where $H_{\rm BH}$ describes the tunnel coupling
and short-range atom-atom interactions,
and $H_{\rm SO}$ and $H_d$ describe
the SO couplings and magnetic
DDIs between the atoms, respectively.

Let us describe this system in more detail.
The two-component Bose-Hubbard model can be used to
describe the interactions of 
two-component bosons in an optical lattice.
We consider this system to have open boundary
conditions. The Hamiltonian $H_{\rm BH}$ can be written as, ($\hbar=1$),
\begin{eqnarray}
H_{\rm BH}&=&\sum_{\alpha,i}\Big[\epsilon^\alpha_i{n^\alpha_i}
+J_\alpha(\alpha^\dag_i\alpha_{i+1}+{\rm H.c.})
+\frac{U_\alpha}{2}{n}^\alpha_i(n^\alpha_i-1)\nonumber\\
&&+U_{ab}n^a_in^b_i\Big],
\end{eqnarray}
where $\alpha_i$ and $\alpha^\dag_i$ are the annihilation
and creation operators of a single atom in the state $|\alpha\rangle$, 
and $n^\alpha_i$ is the number operator at site $i$, and $\alpha=a,b$.
The parameter $\epsilon^\alpha_i$ is the strength of
harmonic confinement at site $i$,
$J_\alpha$ is the tunnel coupling,
$U_{a(b)}$ and $U_{ab}$ are atomic interaction strengths of the intra- and inter-component of atoms, respectively.
For simplicity, we consider the tunnel coupling and atom-atom interaction strength of each component to be nearly equal, i.e., $J_a{\approx}J_b{\approx}J$ and
$U_a{\approx}U_{b}{\approx}{U}$.
We assume that the atoms 
in each component
are additionally trapped in a very shallow harmonic
potential, and therefore $\epsilon^\alpha_i\approx\epsilon^\alpha$
are roughly equal to each other.

Spin-orbit coupling can be generated by inducing
two-photon Raman transition using a pair of lasers \cite{Lin}. By choosing the appropriate phases of lasers, the opposite spin states
of atoms at two neighbouring sites can be coupled \cite{Liu}.  Alternatively, the method for generating SO coupling without
using near resonant light has also been proposed \cite{Kennedy}. The Hamiltonian, which describes SO coupling between two sites, can be written as \cite{Liu,Cai}
\begin{eqnarray}
H_{\rm SO}&=&t_{so}\sum_{i}(a^\dag_ib_{i+1}-a^\dag_ib_{i-1}+{\rm H.c.}),
\end{eqnarray}
where $t_{so}$ is the strength of spin-orbit coupling.

We consider the atoms to interact with
each other due to their magnetic dipoles.
We consider the relative position of 
two magnetic dipoles to be $\vec{r}$.
The DDI energy between the two atoms
is given by \cite{Hensler}
\begin{eqnarray}
U_{dd}(r)&=&\frac{C_{dd}}{4\pi}\frac{(\textbf{S}_1\cdot{\textbf{S}_2})-3(\textbf{S}_1\cdot{\vec{r}})(\textbf{S}_2\cdot{\vec{r}})}{r^3},
\end{eqnarray}
where $C_{dd}=\mu_0(g\mu_B)^2$ is the dipolar interaction strength.
The parameters $\mu_0$, $g$ and $\mu_B$ are the magnetic
permeability of vacuum, the Lande factor, and the Bohr magneton,
respectively. Here $\textbf{S}_{i}$ is the angular momentum operator of the two spin states of atoms.

We consider 
the dipolar interactions which conserve
the angular momentum of internal spin states
between the different sites.
Indeed, the dipolar interactions
also lead to the transition between the different
spin states with accompanying the change of 
orbital angular momentum. By applying a
sufficiently small magnetic field, the
non-conserving terms of atomic spin
states can be neglected \cite{Paz} due to the
large energy difference between the change of
magnetic energy and the potential depth in 
the lattice \cite{Paz0}.

\section{Mott-insulating regime}
We discuss the bosonic lattice in the strong
atom-atom interaction regime, and the lattice
is at unit filling. We assume that the atom-atom
interaction strengths ${U}$ and $U_{ab}$ are repulsive.
We consider the atomic interaction strengths $U$ and $U_{ab}$ to be
much larger than the parameters such as $J$, $t_{so}$ and $|\epsilon^b-\epsilon^a|$. 
In this strongly interacting regime, a single atom 
is allowed to occupy in each potential well. 
We have the four degenerate states for the two neighbouring
sites, i.e.,
$|1^a_i,0^b_i\rangle|0^a_{i+1},1^b_{i+1}\rangle$,
$|1^a_i,0^b_i\rangle|1^a_{i+1},0^b_{i+1}\rangle$,
$|0^a_i,1^b_i\rangle|0^a_{i+1},1^b_{i+1}\rangle$
and
$|0^a_i,1^b_i\rangle|1^a_{i+1},0^b_{i+1}\rangle$.
In this case, the two-mode bosonic operators can be written
in terms of angular momentum operators, i.e.,
$S^+_{i}=a^\dag_ib_i$, $S^-_{i}=b^\dag_i{a_i}$ and $S^z_{i}=(b^\dag_i{b_i}-a^\dag_i{a}_i)/2$.
By using the 
second-order perturbation theory, the effective
Hamiltonian, describes the tunnel and spin-orbit
couplings, is written as \cite{Piraud,Zhao}
\begin{eqnarray}
\label{effsHam}
H^s_{\rm eff}&=&\lambda\sum^{N-1}_{i=1}\Big\{\frac{\Delta}{\lambda}{S^z_i}+2\Big[\Big(\frac{t_{so}}{J}\Big)^2-1\Big]\Big(2\frac{U_{ab}}{U}-1\Big)S^z_iS^z_{i+1}\nonumber\\
&&\!\!+\Big(\frac{t_{so}}{J}\Big)^2(S^+_iS^+_{i+1}+S^{-}_iS^-_{i+1})\!-\!(S^+_iS^-_{i+1}+S^-_{i}S^{+}_{i+1})\nonumber\\
&&-4\frac{U_{ab}}{U}\Big(\frac{t_{so}}{J}\Big)(S^z_iS^x_{i+1}-S^x_iS^z_{i+1})\Big\}
\end{eqnarray}
where $\lambda=2J^2/U_{ab}$ and $\Delta=\epsilon^b-\epsilon^a$.
The last terms in Eq.~(\ref{effsHam}) are known
as the DM interactions \cite{Piraud,Zhao}.
Therefore, this model can be mapped onto the XYZ spin model with the DM interactions in a transverse field.

In addition, the Hamiltonian, describes the dipolar
interactions, can be written as \cite{Paz}
\begin{eqnarray}
H_d&=&\sum_{i,j}\eta_{ij}\Big[S^z_{i}S^z_{j}-\frac{1}{4}(S^+_{i}S^-_{j}+S^{-}_{i}S^+_{j})\Big],
\end{eqnarray}
where $\eta_{ij}$ is the 
dipolar interaction strength between site $i$ and $j$.
The parameter $\eta_{ij}$ can be approximated \cite{Lahaye} by 
$C_{dd}(1-3z^2_{ij}/r^2_{ij})/4\pi{r^3_{ij}}$ being multiplied
by the number densities of the two local sites, where
$r_{ij}$ is the relative position between two dipoles at sites $i$ and $j$ and $z$ is the z-component of $r_{ij}$.
The Hamiltonian $H_d$ describes the Heisenberg interactions in
the spin model \cite{Hensler}. We will consider the
atoms to interact with their 
nearest neighbours with the dipolar interaction strength $\eta_1$.
Here the dipolar interaction strengths are comparable to the effective coupling
strengths in Eq.~(\ref{effsHam}) which is derived 
from the second-order perturbation theory.

\section{Topological phase}
\begin{figure}[ht]
\centering
\includegraphics[height=6.0cm]{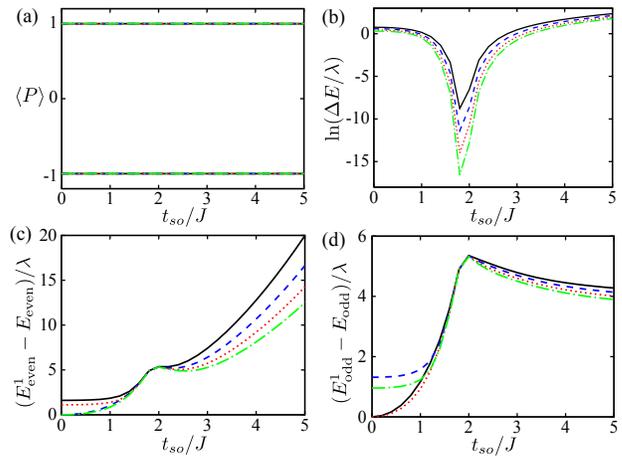}
\caption{ \label{topdeg} (Color online) 
In (a), the expectation of parities $\langle{P}\rangle$ of the two degenerate ground states are plotted.
In (b), logarithmic scale of the energy gap $\Delta{E}$ are plotted versus spin-orbit coupling strength $t_{so}$, for the different $N$s. 
The energy difference between
the first excited and ground states in the even and odd
parities are plotted in (c) and (d), respectively.
The different number of atoms $N$ 
are denoted by the different lines: $N=8$ (black solid),
$N=10$ (blue dashed), $N=12$ (red dotted) and $N=14$ (green dash-dotted). The nearest neighbour dipolar
interaction strength $\eta_1$ is equal to $5\lambda$
and $U_{ab}/U=0.25$.
}
\end{figure}
Since the spin model has the $\mathbb{Z}_2$ 
symmetry, the parity number $P$ is preserved,
where $P$ is defined as \cite{Kitaev2}
\begin{eqnarray}
P&=&\prod^{N}_{i=1}{\sigma^z_i}.
\end{eqnarray}
In this model, there are two parity sectors 
which are called even ($P=1$) and odd ($P=-1$) parities,
respectively.

We consider the energy difference between the
two ground states in the odd and even
parities to indicate the topological phase
of a spin chain \cite{Stoudenmire}. It is defined as 
\begin{eqnarray}
\label{DeltaE}
\Delta{E}&=&|E_{\rm odd}-E_{\rm even}|,
\end{eqnarray}
where $E_{\rm odd}$ and $E_{\rm even}$
are the ground-state energies of the odd-
and even-parity sectors. The system is in
a topological phase if the energy
difference $\Delta{E}$ between two different
parity sectors is zero. Thus, $\Delta{E}$
can act as an order parameter for 
topological phases.

\subsection{Effect of dipolar interactions}
\begin{figure}[ht]
\centering
\includegraphics[height=12.0cm]{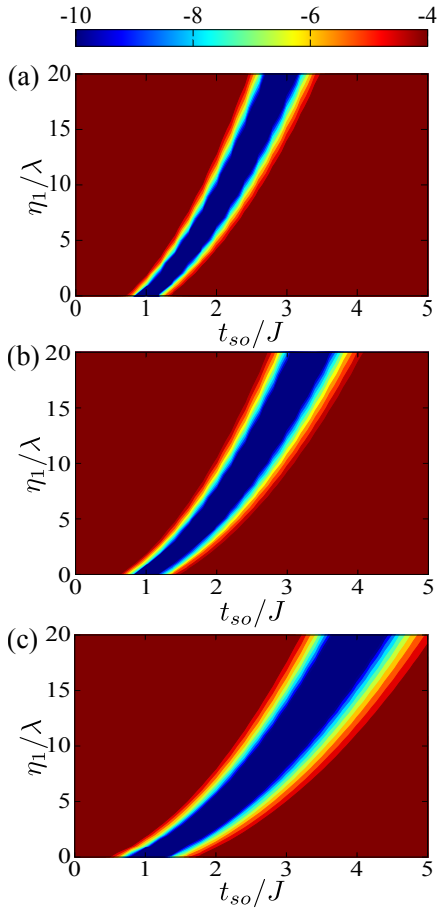}
\caption{ \label{fig_dip_soc} (Color online) Contour plot
of $\ln{(\Delta{E}/\lambda)}$ versus dipolar interaction $\eta_1$
and spin-orbit coupling strength $t_{so}$,
for $N=14$. The different ratios of $U_{ab}/U$ are
shown: (a) $U_{ab}/U=0$, (b) $U_{ab}/U=0.25$ and (c)
$U_{ab}/U=0.5$.  }
\end{figure}

We study the relationship between spin-orbit
couplings and dipolar interactions to the
topological degeneracy. We numerically
solve the eigen-energy 
by using exact diagonalization method.
It should be noted that the DM terms in Eq.~(\ref{effsHam}) break
the $\mathbb{Z}_2$ symmetry. 
We plot the parities of the two degenerate ground
states in Fig.~\ref{topdeg}(a). 
When the ratio $U_{ab}/U$ of the inter- and
intra-component interactions
is smaller than one, this system exhibits the two
definite parities $1$ and $-1$ in Fig.~\ref{topdeg}(a), respectively.
This means that the two degenerate ground
states preserve their parities if
the DM terms are small enough.
To show the topological degeneracy, 
we study the energy gap $\Delta{E}$ in Eq.~(\ref{DeltaE})
versus the SO coupling
in Fig.~\ref{topdeg}(b), for the different sizes of system. 
The energy splitting
decreases exponentially when the size $N$ grows.
This shows the feature of the topological degeneracy \cite{Kitaev2}.

\begin{figure}[ht]
\centering
\includegraphics[height=12.0cm]{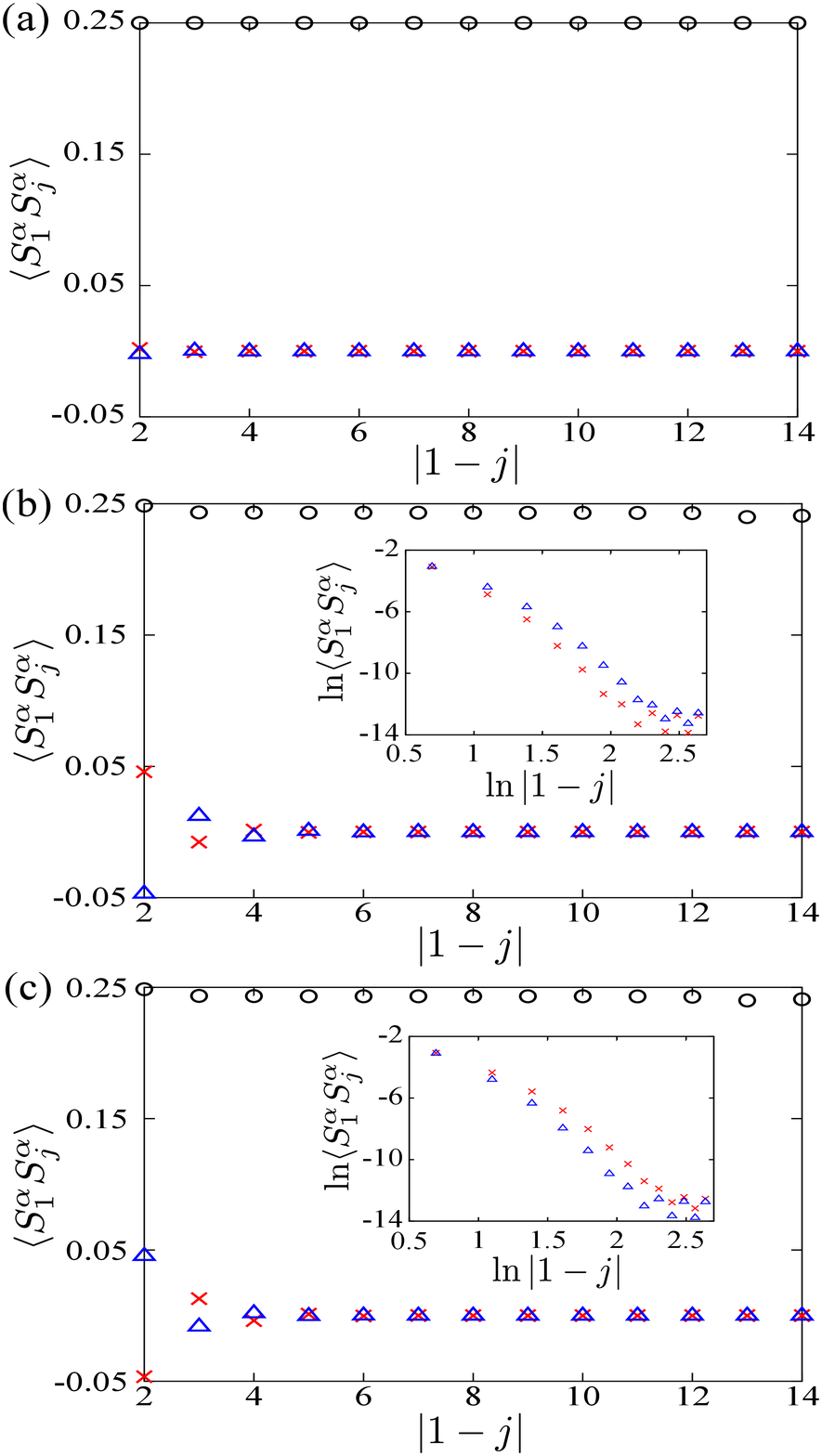}
\caption{ \label{spincorrelation} (Color online) Spin
correlations versus $|1-j|$, for $N=14$ and $U_{ab}/U=0.5$. $\langle{S^x_1}S^x_j\rangle$, $\langle{S^y_1}S^y_j\rangle$ and $\langle{S^z_1}S^z_j\rangle$ are denoted by
red upper-triangle, black circle and blue cross, respectively. The different of SO couling strengths
are shown: (a) $t_{so}=2.9J$, (b) $t_{so}=2.6J$ and
(c) $t_{so}=3.3J$. The two insets in (b) and (c)
show the log-log plots of spin correlations $\langle{S^x_1}S^x_j\rangle$
and $\langle{S^z_1}S^z_j\rangle$ versus the distance $|1-j|$.
}
\end{figure}

We also plot the energy difference between
the first excited and ground states for 
the even- and odd-parity sectors in Fig.~\ref{topdeg}(c)
and (d), respectively. The energy gaps are shown
in the different parities, where they are much 
larger than the energy splitting between 
the two nearly degenerate states.
Therefore, the two degenerate
ground states can be protected by the energy gaps.

\begin{figure}[ht]
\centering
\includegraphics[height=8cm]{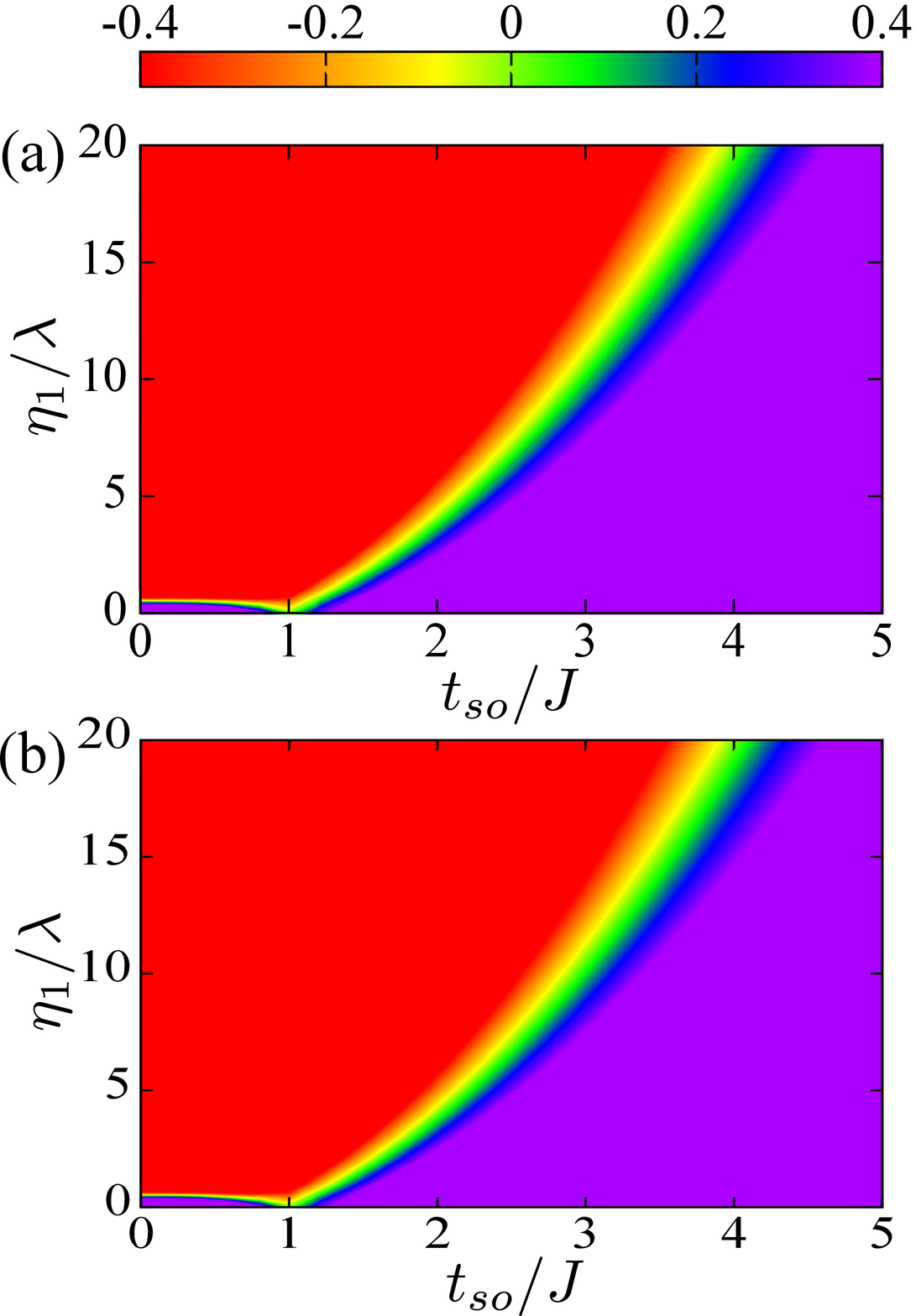}
\caption{ \label{scorr_dip_soc} (Color online) Contour plots
of parameter $S$ versus dipolar interaction $\eta_1$
and spin-orbit coupling strength $t_{so}$,
for $N=14$ and $U_{ab}/U=0.5$. The ground states in even- and odd-parity
are shown in (a) and (b), respectively.}
\end{figure}

In Fig.~\ref{fig_dip_soc}, we plot the contour plot 
of $\ln(\Delta{E})$ versus the dipolar interaction and 
spin-orbit coupling strengths, for the different
ratios of $U_{ab}/U$. The blue area is shown
when the SOC strength increases with the
repulsive dipolar interaction strength. 
This blue area denotes 
that the energy difference $\Delta{E}$
is below $1\times{10^{-4}}$. In addition,
the blue region becomes larger when the dipolar
interaction strength increases as shown in Fig.~\ref{fig_dip_soc}. 
This result
suggests that the repulsive dipolar interactions
can widen the range of SO coupling strengths for 
exhibiting the topological phases. 

Indeed, the topological phase can be described by the ground states of the Ising model along the $y$-direction. When $t_{so}=J$ and $U_{ab}/U=0$, the Hamiltonian $H^s_{\rm eff}$ in Eq.~(\ref{effsHam}) can be reduced to $-4\lambda\sum_{i}S^y_iS^y_{i+1}$.
The two degenerate ground states are $|\!\uparrow\uparrow\ldots\uparrow\rangle_y$ and $|\!\downarrow\downarrow\ldots\downarrow\rangle_y$,
where $|\!\!\uparrow\rangle_j$ and $|\!\!\downarrow\rangle_j$
are spin up and down states in the $y$-direction
at site $j$. ``Tunneling'' between these two 
ground states occurs due to the virtual transitions
via the other perturbation terms in the Hamiltonian \cite{Kitaev2}. Therefore, it falls off exponentially with the size $N$ as shown in Fig.~\ref{topdeg}(b).
When the dipolar interaction increases and
$U_{ab}/U$ is smaller than one, the topological
phase can still be described in this picture, i.e.,
the terms $-\sum_iS^y_iS^y_{i+1}$ are dominant.
This can be manifested in studying the spin correlations.

In Fig.~\ref{spincorrelation}(a), we plot the spin correlations $\langle{S^y_1S^y_j}\rangle$ versus the distance $|1-j|$, where
$U_{ab}/U=0.5$, $t_{so}=2.9J$ and $\alpha=x,y,z$.
Spin correlations $\langle{S^y_1S^y_j}\rangle$ are equal to 0.25 and do not decay between the distant spins, while $\langle{S^x_1S^x_j}\rangle$
and $\langle{S^z_1S^z_j}\rangle$ almost vanish.
We also study the spin correlations near the boundary
of topological phase, where $t_{so}=2.6J$ and $3.3J$. In Figs.~\ref{spincorrelation}(b) and (c),
$\langle{S^y_1S^y_j}\rangle$ is slightly smaller than 0.25 and $\langle{S^x_1S^x_j}\rangle$
and $\langle{S^z_1S^z_j}\rangle$ decay algebraically
in both cases. Therefore, the system starts to show the feature of the incomplete ferromagnet phase \cite{Piraud}.

Next, we study the spin correlations of the two
degenerate ground states. 
We introduce a parameter $S$ which
is defined as
\begin{eqnarray}
\label{scorrelation}
S&=&\sum^{N-1}_{i=1}\langle{S^z_iS^z_{i+1}}\rangle.
\end{eqnarray}
If two spins are in parallel, then the spin
correlation $\langle{S^z_iS^z_{i+1}}\rangle$
will be positive. Otherwise, the two spins
are anti-parallel if $\langle{S^z_iS^z_{i+1}}\rangle$
is negative.
When $S$ is positive(negative), there is
a higher probability to find the two
nearest spins in parallel(anti-parallel).  
If $S$ becomes zero, then there is an
equal probability of finding the two spins
in parallel or anti-parallel. 
Since the topological phase can be
described by the degenerate states $|\!\!\uparrow\uparrow\ldots\uparrow\rangle_y$ and $|\!\!\downarrow\downarrow\ldots\downarrow\rangle_y$, the parameter $S$ becomes zero. Thus, 
this parameter $S$ can thus be used to
indicate the topological phase as long
as $U_{ab}/U$ is smaller than one.

In Fig.~\ref{scorr_dip_soc}, we plot
the contour plots of $S$ versus the 
dipolar interaction and SO coupling
for the two ground states in the even- and
odd-parity in (a) and (b), respectively. 
It is corresponding to Fig.~\ref{fig_dip_soc}(c), where
$U_{ab}/U$ is equal to $0.5$.
In Fig.~\ref{scorr_dip_soc}(a) and (b), 
the left part is red $(S>0)$ and the right part
is purple ($S<0$). The small region, which gives 
$S\approx{0}$, is in between 
these two different regions. This region ($S\approx{0}$) coincides the 
region which shows topological degeneracy
in Fig.~\ref{fig_dip_soc}(c). Therefore, the parameter
$S$ can be used for detecting the topological
phases.

\subsection{Effect of transverse field in the presence of dipolar interactions}
\begin{figure}[ht]
\centering
\includegraphics[height=12cm]{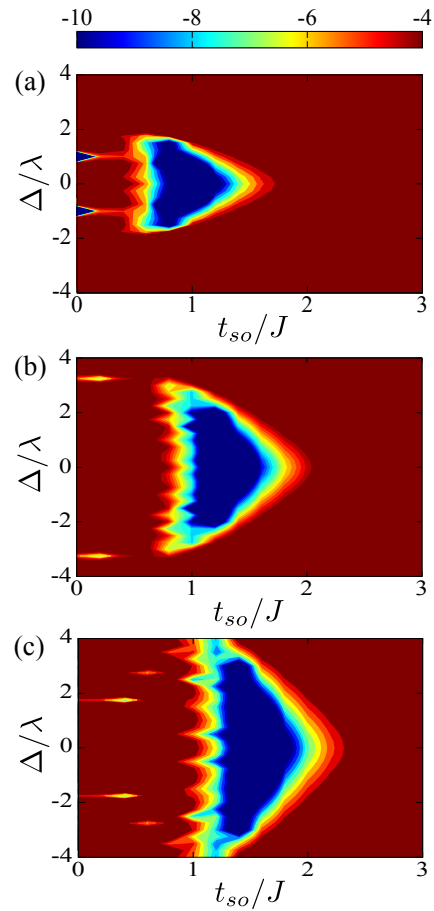}
\caption{ \label{dipolar_field1} (Color online) Contour plots
of $\ln{(\Delta{E}/\lambda)}$ versus transverse
field strength $\Delta$ and spin-orbit coupling strength $t_{so}$ ,
for $N=14$ and $U_{ab}/U=0.5$. The different strengths of dipolar
interaction $\eta_1$ are plotted in (a) $\eta_1=0$,
(b) $\eta_1=\lambda$ and (c) $\eta_1=2\lambda$, respectively.}
\end{figure}
\begin{figure}[ht]
\centering
\includegraphics[height=8cm]{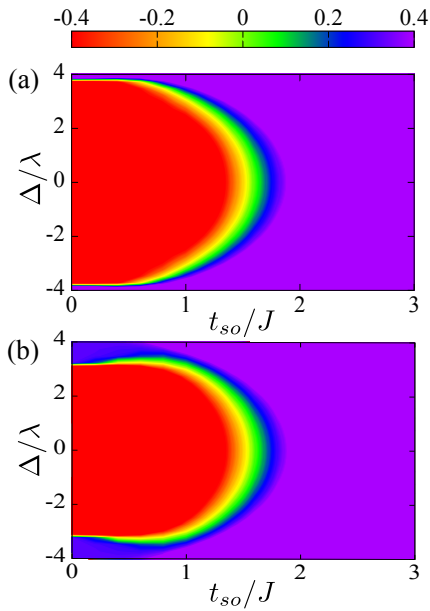}
\caption{ \label{dipfieldscorr} (Color online) Contour plots
of parameter $S$ versus transverse
field strength $\Delta$ and spin-orbit coupling strength $t_{so}$,
for $N=14$, $U_{ab}/U=0.5$ and $\eta_1=2\lambda$. The ground states 
in even- and odd-parity are shown in (a) and (b), respectively.
}
\end{figure}
We study the effect of the transverse field on topological 
degeneracy in the presence of repulsive DDIs.
In Fig.~\ref{dipolar_field1}, the logarithmic scale of
$\Delta{E}$ are plotted versus the transverse field and SO coupling,
for the different dipolar interaction strengths. The stronger 
dipolar interaction strength are shown from Fig.~\ref{dipolar_field1}(a) to (c). We can see that the blue area becomes larger when
the dipolar interactions are stronger. 
This means that the topological
degeneracy can be obtained from a wider range of parameters $\Delta$ and $t_{so}$ when the dipolar interaction strength increases. The repulsive dipolar interactions are therefore useful 
to exhibit the topological phases.

Then, we investigate the parameter S in Eq.~(\ref{scorrelation}) for indicating the topological
phase.
In Fig.~\ref{dipfieldscorr}, the contour plots
of the parameter $S$ are shown versus the transverse
field and SO coupling, for the even- and odd-parity
in (a) and (b), respectively and $\eta_1=2\lambda$. 
The half of an ellipse is 
shown to be red when the parameter $S$ is negative.
The outside region is purple where 
$S$ is positive.
The boundary of the red half ellipse indicates $S$ to be equal to zero.
To compare figures \ref{dipolar_field1}(c)
and \ref{dipfieldscorr}, this small region coincides with the blue area which shows the topological degeneracy in Fig.~\ref{dipolar_field1}(c).  Therefore, the parameter $S$ is
an useful parameter to indicate the topological phase.

\section{Discussion}
Let us estimate the physical parameters
for observing the topological phase. 
It should be noted that the ratio
$J/U$ can be tuned in experiments. As demonstrated in the experiment \cite{Trotzky}, the superexachange interaction strength $2\lambda=4J^2/U_{ab}$ can be adjusted to about $5$ Hz with a high barrier depth. The dipolar coupling strength between the atoms in the nearest site is about $\sim~20$ Hz if chromium atoms are used. This has been shown in the recent experiment \cite{Paz}. Therefore, the ratio of the dipolar interaction strength $\eta_1$ ($\sim{20}$ Hz) to the superexchange interaction strength $\lambda$ ($\sim{2}$ Hz) can attain $\sim{10}$. Therefore, the effect of dipolar interactions on the topological phase can be detected with the current experimental setting.

\section{Conclusion}
In summary, we have studied 
a 1D spin-orbit coupled dipolar
lattice bosons, where the atoms
at the different sites can be 
coupled via the magnetic dipolar
interactions. In the Mott-insulating
regime, this system can be described
by the quantum XYZ model with the DM 
interactions in a transverse
field. We adopt the energy difference
between the two ground states in the
different parities to determine the topological
phase. The repulsive DDI can broaden the parameter range of SO coupling and transverse field strengths
for exhibiting topological phases. We show that
the sum of the spin polarization correlations between
two nearest neighbouring atoms can be used to indicate 
topological phases. It may be useful to
detect the topological phase in experiments.

\begin{acknowledgments}
We thank Jize Zhao and Marie Piraud for useful
discussion.
This work was supported in part by the 
National Basic Research Program of 
China Grants No. 2011CBA00300 and No. 2011CBA00301 
the National Natural Science Foundation of 
China Grants No. 11304178, No. 61061130540, and No.
61361136003.  
\end{acknowledgments}

\end{document}